\documentclass[12pt,epsf]{article}
\usepackage{amssymb,amsmath}
\usepackage{graphicx}
\usepackage{color}
\usepackage[hidelinks]{hyperref}

\parskip=\baselineskip

\newcommand{\be}{\begin{equation}}
\newcommand{\ee}{\end{equation}}
\newcommand{\bea}{\begin{eqnarray}}
\newcommand{\eea}{\end{eqnarray}}
\newcommand{\beas}{\begin{eqnarray*}}
\newcommand{\eeas}{\end{eqnarray*}}
\newcommand{\ba}{\begin{array}}
\newcommand{\ea}{\end{array}}
\newcommand{\tr}{{\rm tr}}

\newcommand{\cDsl}{{{\cal D}\kern-.65em /\,}}
\newcommand{\cDslsm}{{{\cal D}\kern-.5em /\,}}
\newcommand{\nabslsm}{\nabla\kern-.55em /}
\newcommand{\pasl}{\pa\kern-.55em /}
\newcommand{\psl}{p\kern-.45em /}
\newcommand{\Dsl}{D\kern-.65em /}
\newcommand{\Asl}{A\kern-.55em /}
\newcommand{\nabsl}{\nabla\kern-.65em /\kern+.2em}
\newcommand{\qsl}{q\kern-.5em /}
\newcommand{\ksl}{k\kern-.5em /}
\newcommand{\rsl}{r\kern-.5em /}

\newcommand{\cDslLCsq}{{\stackrel{\circ}{\cDsl^{\kern2pt 2}}}}

\newcommand\cc[1]{#1^{^{\kern-6pt \circ}}\kern2pt}

\newcommand{\la}{\langle}
\newcommand{\ra}{\rangle}

\renewcommand{\d}{\delta}

\def\tr{{\rm tr}}

\newcommand{\pa}{\partial}

\newcommand{\beq}{\begin{equation}}
\newcommand{\eeq}{\end{equation}}
\newcommand{\beqn}{\begin{eqnarray}}
\newcommand{\eeqn}{\end{eqnarray}}

\def\dalemb#1#2{{\vbox{\hrule height .#2pt
\hbox{\vrule width.#2pt height#1pt \kern#1pt
\vrule width.#2pt}
\hrule height.#2pt}}}

\begin{document}

\begin{titlepage}
\hfill
\vbox{
    \halign{#\hfil         \cr
           } 
      }  
\vspace*{0mm}
\begin{center}
{\Large \bf Positive gravitational subsystem energies \\  from CFT cone relative entropies}

\vspace*{10mm}
\vspace*{1mm}
Dominik Neuenfeld, Krishan Saraswat, and Mark Van Raamsdonk
\vspace*{1cm}
\let\thefootnote\relax\footnote{dneuenfe@phas.ubc.ca, ksaraswa@ualberta.ca, mav@phas.ubc.ca}

{Department of Physics and Astronomy,
University of British Columbia\\
6224 Agricultural Road,
Vancouver, B.C., V6T 1W9, Canada}

\vspace*{1cm}
\end{center}

\begin{abstract}
The positivity of relative entropy for spatial subsystems in a holographic CFT implies the positivity of certain quantities in the dual gravitational theory. In this note, we consider CFT subsystems whose boundaries lie on the lightcone of a point $p$. We show that the positive gravitational quantity which corresponds to the relative entropy for such a subsystem $A$ is a novel notion of energy associated with a gravitational subsystem bounded by the minimal area extremal surface $\tilde{A}$ associated with $A$ and by the AdS boundary region $\hat{A}$ corresponding to the part of the lightcone from $p$ bounded by $\partial A$. This generalizes the results of arXiv:1605.01075 for ball-shaped regions by making use of the recent results in arXiv:1703.10656 for the vacuum modular Hamiltonian of regions bounded on lightcones. As part of our analysis, we give an analytic expression for the extremal surface in pure AdS associated with any such region $A$. We note that its form immediately implies the Markov property of the CFT vacuum (saturation of strong subadditivity) for regions bounded on the same lightcone. This gives a holographic proof of the result proven for general CFTs in arXiv:1703.10656. A similar holographic proof shows the Markov property for regions bounded on a lightsheet for non-conformal holographic theories defined by relevant perturbations of a CFT.
\end{abstract}


\end{titlepage}
\section{Introduction}

Via the AdS/CFT correspondence, it is believed that any consistent quantum theory of gravity defined for asymptotically AdS spacetimes with some fixed boundary geometry ${\cal B}$ corresponds to a dual conformal field theory defined on ${\cal B}$. Recently, it has been understood that many natural quantum information theoretic quantities in the CFT correspond to natural gravitational observables (see, for example \cite{Ryu:2006bv}, or \cite{VanRaamsdonk:2016exw,Rangamani:2016dms} for a review). Through this correspondence, properties which hold true for the quantum information theoretic quantities can be translated to statements about gravitational physics. In this way, we can obtain a alternative/deeper understanding of some known properties gravitational systems, but also discover novel properties that must hold in consistent theories of gravity.
A particularly interesting quantum information theoretic quantity to consider is relative entropy \cite{Blanco:2013joa}. For a general state $|\Psi \rangle$ of the CFT, we can associate a reduced density matrix $\rho_A$ to a spatial region $A$ by tracing out the degrees of freedom outside of $A$. Relative entropy $S(\rho_A || \rho_A^0)$, which we review in section \ref{sec:background}, quantifies how different this state is from the vacuum density matrix $\rho_A^0$ reduced on the same region. Relative entropy is typically UV-finite, always positive, and has the property that it increases as we increase the size of the region $A$ (known as the monotonicity property). According to the AdS/CFT correspondence, this should correspond to some quantity in the gravitational theory which also obeys these positivity and monotonicity properties. This has previously been explored in \cite{Banerjee:2014oaa, Banerjee:2014ozp,Lin:2014hva, Lashkari:2014kda,Bhattacharya:2014vja, Lashkari:2015hha,Lashkari:2016idm}.

As we review in section \ref{sec:background}, by making use of the holographic formula relating CFT entanglement entropies to bulk extremal surface areas (the ``HRRT formula'' \cite{Ryu:2006bv, Hubeny:2007xt}), it is possible to explicitly write down the gravitational quantity corresponding to relative entropy as long as the vacuum modular Hamiltonian ($H_A^0 = -\log \rho_A^0$) for the region $A$ is ``local'', that is, it can be written as a linear combination of local operators in the CFT. Until recently, such a local form was only known for the modular Hamiltonian of ball-shaped regions \cite{Casini:2011kv}. For these regions, relative entropy has been shown to correspond to an energy that can be associated with the bulk entanglement wedge corresponding to this ball \cite{Lashkari:2014kda,Lashkari:2015hha}.\footnote{The entanglement wedge is a region defined by the union of spacelike surfaces with one boundary on the HRRT surface from the ball and the other boundary on the domain of dependence of the ball at the AdS boundary.} The positivity of relative entropy then implies an infinite family of positive energy constraints (reviewed below) \cite{Lashkari:2016idm}.

Ball-shaped regions (of Minkowski space) have the property that their boundary lies on the past lightcone of a point $p$ and the future lightcone of some other point $q$. In the recent work \cite{Casini:2017roe}, it has been shown that the vacuum modular Hamiltonian for a region $A$ has a local expression so long as the boundary $\partial A$ of $A$ lies on the past lightcone of a point $p$ {\it or} the future lightcone of a point $q$. Thus, we have a much more general class of regions for which the relative entropy and its properties can be interpreted gravitationally. The main goal of the present paper is to explain this interpretation.
\begin{figure}
\centering
\includegraphics[width=0.5\textwidth]{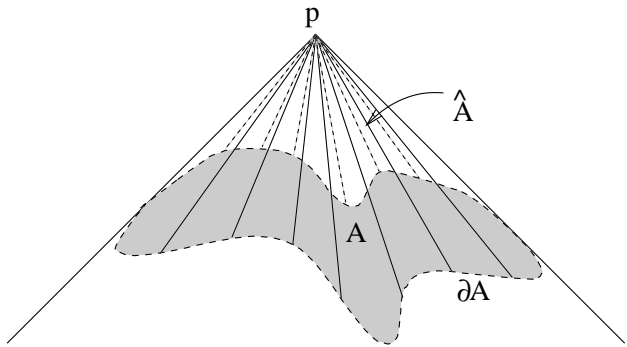}
\caption{Region $A$ whose boundary $\partial A$ is on the past lightcone of a point. }
\label{coneregion}
\end{figure}

In the general case, we denote by $\hat{A}$ the region of the lightcone bounded by $\partial A$, as shown in figure 1. The modular Hamiltonian can then be written as
\be
\label{modH}
H^0_A = \int_{\hat{A}} \zeta_A^\mu(x) T_{\mu \nu}(x) \epsilon^\nu \; ,
\ee
where $T_{\mu \nu}$ is the CFT stress-energy tensor, $\epsilon^\mu$ is a volume form defined in section \ref{sec:background}, and $\zeta_A^\mu(x)$ is a vector field on $\hat{A}$ directed towards the tip of the cone and vanishing at the tip of the cone and on $\partial A$.

\begin{figure}
\centering
\includegraphics[width=0.25\textwidth]{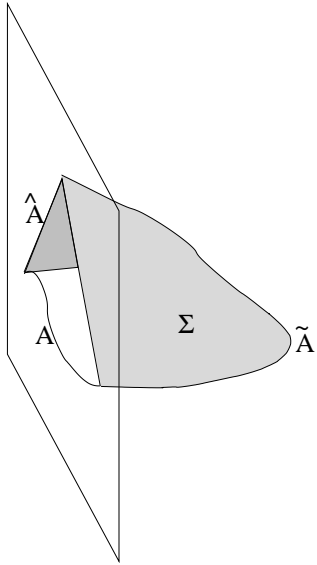}
\caption{CFT Relative entropy associated with boundary region $A$ corresponds to a certain energy associated with a gravitational subsystem defined by the domain of dependence of any spatial region $\Sigma$ bounded by cone region $\hat{A}$ with   $\partial \hat{A} = \partial A$ and extremal surface $\tilde{A}$ with $\partial \tilde{A} = \partial A$. }
\label{adscone}
\end{figure}

To describe the gravitational interpretation of the relative entropy for region $A$, we consider any codimension one spacelike surface $\Sigma$ in the dual geometry such that $\Sigma$ intersects the AdS boundary at $\hat{A}$ and is bounded in the bulk by the HRRT surface $\tilde{A}$ (the minimal area extremal surface homologous to $A$). This is illustrated in figure 2.  Next, we define a timelike vector field $\xi$ in a neighborhood of $\Sigma$ with the properties that $\xi$ approaches $\zeta_A$ at the AdS boundary and behaves near the extremal surface $\tilde{A}$ like a Killing vector associated with the local Rindler horizon at $\tilde{A}$. The timelike vector field $\xi$ represents a particular choice of time on the surface $\Sigma$ and we can define an energy $H_\xi$ associated with this. While generally there are many choices for the surface $\Sigma$ and the vector field $\xi$, we can show that all of them lead to the same value for the energy $H_\xi$. It is this quantity that corresponds to the CFT relative entropy $S(\rho_A || \rho_A^0)$.\footnote{In this paper, we focus on the leading contribution to the CFT relative entropy at large $N$ and make use of the classical HRRT formula. More generally, we expect that the bulk quantity will be corrected by a term $-\Delta S_\Sigma$ measuring the vacuum-subtracted bulk entanglement of the region $\Sigma$.}

The independence of $H_\xi$ on the surface $\Sigma$ used to define it can be understood as a bulk conservation law for this notion of energy. In the case of a ball-shaped region \cite{Lashkari:2016idm}, the energy $H_\xi$ is conserved in a stronger sense (or a bigger volume), since there we are also free to vary the boundary surface $\hat{A}$ to be any spatial surface $A'$ homologous to $A$ in the domain of dependence $D_A$ of $A$. In that case, the vector field $\zeta_A$ can be defined everywhere in $D_A$ such that the expression \eqref{modH} for the modular Hamiltonian gives the same result for any surface $A'$. The bulk vector field $\xi$ can be defined on the full entanglement wedge for $A$, i.e. the union of spacelike surfaces ending on $\tilde{A}$ and on any $A'$ in $D_A$, so we can think of the energy $H_\xi$ as being associated with the entire entanglement wedge. In the more general case considered in this paper, the collection of allowed surfaces $\Sigma$ generally still define a codimension zero region $W_A$ of the bulk spacetime (equivalent to the bulk domain of dependence of any particular $\Sigma$), but this region intersects the boundary only on the lightlike surface $\hat{A}$ rather than the whole domain of dependence region $D_A$.

In section 4, we consider the limit where the geometry is a small deformation away from pure AdS. For pure AdS, we show that the extremal surface $\tilde{A}$ associated with a region $A$ whose boundary lies on the lightcone of $p$ always lies on the bulk lightcone of $p$. Thus, in a limit where perturbations to AdS become small, the wedge $W_A$ collapses to the portion $\hat{A}_{bulk}$ of this lightcone between $p$ and $\tilde{A}$. We present an analytic expression for the extremal surface $\tilde{A}$ and a canonical choice for the vector field $\xi$ on $\hat{A}_{bulk}$. In terms of these, we can write an explicit expression for the leading perturbative contribution to the energy $H_\xi$, which takes the form of an integral over $\hat{A}_{bulk}$ quadratic in the bulk field perturbations.

In section 5, we point out that the explicit form of the extremal surface $\tilde{A}$ in the pure AdS case (in particular, the fact that it lies on the bulk lightcone) leads immediately to a holographic proof of the Markov property for subregions of a CFT in its vacuum state, namely that for two regions $A$ and $B$ the strong subadditivity inequality
\be
S(A)+S(B)-S(A \cap B) - S(A \cup B) \geq 0,
\ee
is saturated if their boundaries lie on the past or future lightcone of the same point $p$. This was shown for general CFTs in \cite{Casini:2017roe}, so it had to hold in this holographic case. The holographic proof extends easily to cases where the field theory is Lorentz-invariant but non-conformal, for example a CFT deformed by a relevant perturbation. In this case, the statement holds for subregions $A$, $B$ whose boundaries lie on a null-plane.

We conclude in section 6 with a discussion of some possible future directions.

\section{Background}
\label{sec:background}

\subsection{Relative entropy in conformal field theories}

For a general quantum system or subsystem described by a density matrix $\rho$, the relative entropy quantifies the difference between $\rho$ and a reference state $\sigma$. It is defined as
\be
S(\rho||\sigma) = \tr(\rho \log \rho) - \tr(\rho \log \sigma) \, ,
\ee
which can be shown to be nonnegative as well as vanishing if and only if $\rho = \sigma$.

Relative entropy also obeys a monotonicity property: when $A$ is a subsystem of the original system, the relative entropy satisfies
\be
\label{monotonicity}
S(\rho_A||\sigma_A) \le S(\rho||\sigma) \, ,
\ee
where $\rho_A$ and $\sigma_A$ are the reduced density matrices for the subsystem defined from $\rho$ and $\sigma$ respectively.

Using the definition $S(\rho) = - \tr(\rho \log \rho)$ of entanglement entropy and $H_\sigma = - \log(\sigma)$ of the modular Hamiltonian associated with $\sigma$, we can rewrite the expression for relative entropy as \cite{Blanco:2013joa}
\be
\label{REdiff}
S(\rho\|\sigma) = \Delta \la H_\sigma\ra - \Delta S
\ee
where $\Delta$ indicates a quantity calculated in the state $\rho$ minus the same quantity calculated in the reference state $\sigma$.

For a conformal field theory in the vacuum state, the modular Hamiltonian of a ball-shaped region takes a simple form \cite{Casini:2011kv}. For a ball $B$ of radius $R$ centered at $x_0$ in the spatial slice perpendicular to the unit timelike vector $u^\mu$, the modular Hamiltonian is
\be
\label{defE}
H_B = \int_{B'} \zeta_B^\mu T_{\mu \nu} \epsilon^\nu,
\ee
where $\epsilon_\nu = \frac{1}{(d-1)!} \epsilon_{\nu \mu_1 \cdots \mu_{d-1}} dx^{\mu_1} \wedge \dots \wedge dx^{\mu_{d-1}}$ is a volume form and $\zeta_B$ is the conformal Killing vector
\be
\label{defzeta}
\zeta_B^\mu = \frac{\pi}{R} \left\{ [R^2 - (x-x_0)^2] u^\mu + [2 u_\nu (x - x_0)^\nu ](x - x_0)^\mu \right\} \, ,
\ee
with some four-velocity $u^\mu$. The modular Hamiltonian is the same for any surface $B'$ with the same domain of dependence as $B$.

Using the expression (\ref{defE}) in (\ref{REdiff}), the relative entropy for a state $\rho$ compared with the vacuum state may be expressed entirely in terms of the entanglement entropy and the stress tensor expectation value. For a holographic theory in a state with a classical gravity dual, these quantities can be translated into gravitational language using the HRRT formula (which also implies the usual holographic relation between the CFT stress-energy tensor expectation value and the asymptotic bulk metric \cite{Faulkner:2013ica}). Thus, the CFT relative entropy for a ball-shaped region corresponds to some geometrical quantity in the gravitational theory with positivity and monotonicity properties. In \cite{Lashkari:2015hha} and \cite{Lashkari:2016idm}, this quantity was shown to have the interpretation of an energy associated with the gravitational subsystem associated with the interior of the entanglement wedge associated with the ball.

Recently, Casini, Test\'e, and Torroba have provided an explicit expression for the vacuum modular Hamiltonian of any spatial region $A$ whose boundary lies on the lightcone of a point \cite{Casini:2017roe}. To describe this, consider the case where $\partial A$ lies on the past lightcone of a point $p$ and let $\hat{A}$ be the region on the lightcone that forms the future boundary of the domain of dependence of $A$. For $x \in \hat{A}$, define a function $f(x)$ that represents what fraction of the way $x$ is along the lightlike geodesic from $p$ through $x$ to $\partial A$ (so that $f(p) = 0$ and $f(x) = 1$ for $x \in \partial A$). Now, define a lightlike vector field on $\hat{A}$ by
\be
\label{defZetaA}
\zeta_A^\mu(x) \equiv  2 \pi (f(x) - 1) (p^\mu - x^\mu) \; .
\ee
Then the modular Hamiltonian can be expressed as
\be
\label{defEA}
H_A = \int_{\hat{A}} \zeta_A^\mu T_{\mu \nu} \epsilon^\nu \; .
\ee
In general, we cannot extend the vector field away from the surface $\hat{A}$ such that the expression (\ref{defEA}) remains valid when integrated over an arbitrary surface $A'$ with $\partial A' = \partial A$. In equation \eqref{HA} we give an explicit expression for $H_A$ in a convenient coordinate frame.

Using this expression in \eqref{REdiff}, we can express the relative entropy for the region $A$ in a form that can be translated to a geometrical quantity using the HRRT formula. We would again like to understand the gravitational interpretation for this positive quantity.

\subsection{Gravity background}

We now focus on states in a holographic CFT dual to some asymptotically AdS spacetime with a good classical description. For any spatial subsystem $A$ of the CFT, there is a corresponding region on the boundary of the dual spacetime (which we will also call $A$). The HRRT formula asserts that the CFT entanglement entropy for the spatial subsystem $A$ in a state $|\Psi \rangle$, at leading order in the $1/N$ expansion, is equal to $1/(4 G_N)$ times the area of the minimal area extremal surface $\tilde{A}$ in the dual spacetime which is homologous to the region $A$ on the boundary.

For pure AdS, when the CFT region is a ball $B$, the spatial region $\Sigma$ between $B$ and $\tilde{B}$ forms a natural ``subsystem'' of the gravitational system, in that there exists a timelike Killing vector $\xi_B$ defined on the domain of dependence $D_\Sigma$ of $\Sigma$ and vanishing on $\tilde{B}$. At the boundary of AdS, this reduces to the vector $\zeta_B$ appearing in the modular Hamiltonian (\ref{defE}) for $B$. The vector $\xi_B$ gives a notion of time evolution which is confined to $D_\Sigma$. From the CFT point of view, this time evolution corresponds to evolution by the modular Hamiltonian (\ref{defE}) within the domain of dependence of $B$, which by a conformal transformation can be mapped to hyperbolic space times time.

For states which are small perturbations to the CFT vacuum state, it was shown in \cite{Lashkari:2015hha} that the relative entropy for a ball $B$ at second order in perturbations to the vacuum state corresponds to the perturbative bulk energy associated with the timelike Killing vector $\xi_B$ in $D_\Sigma$ (known as the ``canonical energy'' associated with this vector).

This result was extended to general states in \cite{Lashkari:2016idm}. While there are no Killing vectors for general asymptotically AdS geometries, it is always possible to define a vector field $\xi_B$ that behaves near the AdS boundary and near the extremal surface in a similar way to the behavior of the Killing vector $\xi_B$ in pure AdS. Specifically, we impose conditions
\bea
\xi^a|_{B} &=& \zeta^a_B, \label{x3} \\
\nabla^{[a}\xi^{b]}|_{\tilde B} &=& 2\pi n^{ab}, \label{x1} \\
\xi|_{\tilde B} &=& 0 \label{x2} \; ,
\label{xicond}
\eea
where $n^{ab}$ is the binormal to the codimension two extremal surface $\tilde{B}$. Given any such vector field, we can define a diffeomorphism
\be
\label{xiflow}
g \to g + {\cal L}_\xi g \; .
\ee
This represents a symmetry of the gravitational theory, so we can define a corresponding conserved current and Noether charge. The resulting charge $H_\xi$ turns out to be the same for any vector field satisfying the conditions \eqref{x3} -- \eqref{x2}. It can be interpreted as an energy associated to the vector field $\xi_B$ or alternatively as the Hamiltonian that generates the flow (\ref{xiflow}) in the phase space formulation of gravity. The main result of \cite{Lashkari:2016idm} is that the CFT relative entropy for a state $|\Psi \rangle$ comparing the reduced density matrix $\rho_B$ to its vacuum counterpart $\rho^{(vac)}_B$ is equal to the difference of this gravitational energy between the spacetime $M_\psi$ dual to $|\Psi \rangle$ and pure AdS,
\be
\label{diffH}
S(\rho_B||\rho^{(vac)}_B) = H_{\xi}(M_\psi) - H_{\xi}(AdS) \; .
\ee
We will review the derivation of this identity in the next section when we generalize it to our case.

To write $H_\xi$ explicitly, we start with the Noether current (expressed as a $d$-form)
\be
\label{JefD}
J_\xi = \theta ({\cal L}_\xi g) - \xi \cdot L \; ,
\ee
where $L$ is the Lagrangian density and $\theta$ is defined by
\be
\label{eom}
\delta
L(g) = d \theta(\delta g) + E(g)\delta g \; .
\ee
Here, $E(g)$ are the equations of motion obtained in the usual way by varying the action. The Noether current is conserved off-shell for Killing vector fields and on-shell for any vector field $\xi$,
\begin{align}
dJ_\xi = E(g) \cdot \mathcal L_\xi g.
\end{align}
Then, up to a boundary term, the energy $H_\xi$ is defined in the usual way as the integral of the Noether charge over a spatial surface:
\be
 \label{energy}
  H_\xi = \int_\Sigma J_\xi - \int_{\partial \Sigma} \xi \cdot K .
\ee
Here, $\Sigma$ is any spacelike surface bounded by the HRRT surface $\tilde{B}$ and by a spacelike surface $\Sigma_{\partial M}$ on the AdS boundary with the same domain of dependence as $B$. For a ball-shaped region $B$, the quantity $H_\xi$ is independent of both the bulk surface $\Sigma$ (as a consequence of diffeomorphism invariance) and also the spacelike surface $\Sigma_{\partial M}$ at the boundary of AdS (as a consequence of the fact that $\zeta_B$ defines an asymptotic symmetry).

The quantity $K$ in the boundary term is defined so that
\be
 \label{integrate_theta} \delta (\xi \cdot K) = \xi \cdot \theta(\delta g) ~~~
 {\rm on \ \partial \Sigma}\, \; .
 \ee
As explained in  \cite{Lashkari:2016idm}, this ensures that the difference (\ref{diffH}) does not depend on the regularization procedure used to calculated the energies and perform the subtraction.

We can rewrite $H_\xi$ completely as a boundary term using the fact that on-shell, $J_\xi$ can be expressed as an exact form \cite{Lashkari:2016idm}
\be
J_\xi = d Q_\xi \; .
\ee
Thus, for a background satisfying the gravitational equations, we have
\be
 \label{energyA}
  H_\xi = \int_{\partial \Sigma} Q_\xi - \int_{\partial \Sigma} \xi \cdot K  \; .
\ee
This shows that the definition of $H_\xi$ is independent of the details of the vector field $\xi$ in the interior of $\Sigma$. In our derivations below, it will be useful to have a differential version of this expression that gives the change in $H_\xi$ under on-shell variation of the metric. By combining (\ref{energyA}) with (\ref{integrate_theta}), we obtain
\be
 \label{energyDiff}
  \delta H_\xi = \int_{\partial \Sigma} (\delta Q_\xi - \xi \cdot \theta)  \;
\ee

The interpretation  of $H_\xi$ as a Hamiltonian for the phase space transformation associated with (\ref{xiflow}) can be understood by recalling that the symplectic form on this phase space is defined by
\be
\Omega(\delta g_1, \delta g_1) = \int_\Sigma \omega(g,\delta g_1, \delta g_1)
\ee
where the $d$-form $\omega$ is defined in terms of $\theta$ as
\begin{align}
\omega(g, \delta_1 g, \delta_2 g) = \delta_1 \theta(g, \delta_2 g) - \delta_2 \theta(g,\delta_1 g) \; .
\end{align}
In terms of $\omega$ we have that for an arbitrary on-shell metric perturbation
\be
\label{eq:deltaH}
\delta H_\xi = \Omega(\delta g,{\cal L}_\xi g) = \int_\Sigma \omega(g, \delta g, {\cal L}_\xi g)
\ee
This amounts to the usual relation $d H = v_H \cdot \Omega$ between a Hamiltonian (in this case $H_\xi$) and its corresponding vector field (in this case ${\cal L}_\xi g$) via the symplectic form $\Omega$.

\section{Bulk interpretation of relative entropy for general regions bounded on a lightcone}
\label{sec:BulkInterpretation}

Consider now a more general spacelike CFT subsystem $A$ whose boundary lies on some lightcone. In this case -- unless the boundary is a sphere -- there is no longer a conformal Killing vector defined on the domain of dependence region $D_A$ and we cannot write the boundary modular Hamiltonian as in (\ref{defE}) where the result is independent of the surface $\hat{B}$. Nevertheless, we have a similar expression (\ref{defEA}) for the modular Hamiltonian as a weighted integral of the CFT stress tensor over the lightcone region $\hat{A}$ (shown in Figure 1). Thus, making use of the formula (\ref{REdiff}) for relative entropy, together with the holographic entanglement entropy formula and the holographic dictionary for the stress-energy tensor, we can translate the CFT relative entropy to a gravitational quantity. In this section, we show that this can again be interpreted as an energy difference,
\be
S(\rho_A||\rho^{vac}_A) = H_\xi(M_\psi) - H_\xi(AdS)
\ee
for an energy $H_\xi$ associated with a bulk spatial region $\Sigma$ bounded by $\hat{A}$ and the bulk extremal surface $\tilde{A}$.

To begin, we choose a bulk vector field $\xi$ satisfying
\bea
\xi^a|_{\hat{A}} &=& \zeta^a_A, \label{A3} \\
\nabla^{[a}\xi^{b]}|_{\tilde A} &=& 2\pi n^{ab}, \label{A1} \\
\xi|_{\tilde A} &=& 0, \label{A2}
\eea
The argument that the latter two conditions can be satisfied is the same as in \cite{Lashkari:2016idm}, making use of the fact that we can define Gaussian null coordinates near the surface $\tilde{A}$. To enforce the first condition, we will make use of Fefferman-Graham (FG) coordinates for which the near-boundary metric takes the form
\be
ds^2 = \frac{1}{z^2} (dz^2 + dx_\mu dx^\mu + z^d \Gamma_{\mu \nu}^{(d)} dx^\mu dx^\nu + {\cal O}(z^{d+1}))
\ee
and choose a vector field expressed in these coordinates as
\bea
\label{eq:behaviorVectorField}
\xi^\mu &=& \zeta_A^\mu + z \xi^\mu_1 + z^2 \xi^\mu_2 + \dots \cr
\xi^z &=& z \xi^z_1 + z^2 \xi^z_2 + \dots \; .
\eea

We will now evaluate $\delta H_\xi$ for this vector field starting from (\ref{energyDiff}) and find that it matches with a holographic expression for the change in relative entropy. First, we evaluate the part at the AdS boundary. Explicit calculations in the FG gauge, which are done in appendix \ref{app:modularHamiltonian}, show that
\be
\delta Q_\xi - \xi \cdot \theta|_{z \to 0} =\frac{d}{16\pi G_N} \xi^a \delta \Gamma_{ab}^{(d)} \hat \epsilon^b|_{z \to 0}=\frac{d}{16\pi G_N} \zeta^{\mu}_{\hat{A}} \delta \Gamma_{\mu\nu}^{(d)} {\epsilon}^{\mu} \; ,
\ee
where $\epsilon$ was defined in the previous section and
\begin{align}
\hat \epsilon_{a_1 \dots a_k} =  \frac{\sqrt{-g}}{(d+1-k)!} \epsilon_{a_1 \dots a_k b_1 \cdots b_{d+1-k}} dx^{b_1} \wedge \dots \wedge dx^{b_{d+1-k}}.
\end{align}
Using the standard holographic relation between the asymptotic metric and the CFT stress tensor expectation value, we obtain
\be
\int_{\hat{A}} (\delta Q_\xi - \xi \cdot \theta) = \frac{d}{16 \pi G_N} \int_{\hat{A}} \zeta_{\hat{A}}^\mu \delta \Gamma_{\mu \nu}^{(d)} {\epsilon}^\mu = \int_{\hat{A}} \zeta^{\mu}_{\hat{A}} \delta \left<T_{\mu\nu}\right>{\epsilon}^{\mu} = \delta\left<H_{\hat{A}}\right> \;.
\ee
Here, $H_{\hat{A}}$ is the boundary modular Hamiltonian for the region $A$, so this term represents the variation in the modular Hamiltonian term in the expression (\ref{REdiff}) for relative entropy.

Next, we look at the part of (\ref{energyDiff}) coming from the other boundary of $\Sigma$, at the extremal surface. By condition \eqref{A2} we have that $\xi$ vanishes on $\tilde A$ and we are left with the integral over $\delta Q_\xi$.
$Q_\xi$ can be brought into the form $\frac 1 {16\pi} \nabla^a \xi^b \hat \epsilon_{ab}$ \cite{Hollands:2012sf} and by virtue of \eqref{A1} we obtain the entanglement entropy using the HRRT conjecture,
\be
\int_{\tilde A} \delta Q_\xi = \frac 1 {4  G_N} \int_{\tilde A} = \delta S.
\ee

Combining both contributions to (\ref{energyDiff}), we have that
\be
\delta H_\xi = \delta\left<H_{\hat{A}}\right> - \delta S,
\ee
where the variation corresponds to an infinitesimal variation of the CFT state. Integrating this from the CFT vacuum state up to the state $|\psi \rangle$, we have that
\be
S(\rho_A||\rho_A^{vac}) = \Delta\left<H_{\hat{A}}\right> - \Delta S =  \Delta H_\xi
\ee
Thus, we have established that for a boundary region $A$ with $\partial A$ on a lightcone, the CFT relative entropy is interpreted in the dual gravity theory as an energy associated with the timelike vector field $\xi$.

The energy $H_\xi$ is naturally associated with a certain spacetime region of the bulk, foliated by spatial surfaces bounded by the boundary lightcone region $\hat{A}$ and the bulk extremal surface $\tilde{A}$. That such spatial surfaces exist is a consequence of the fact that the extremal surface $\tilde{A}$ always lies outside the causal wedge of the region $A$ (the intersection of the causal past and the causal future of the domain of dependence of $A$) \cite{Wall:2012uf}.

\section{Perturbative expansion of the holographic dual to relative entropy}
\label{sec:perturbations}
In this section, we consider the expression for $H_\xi$ in the case where the CFT state is a small perturbation of the vacuum state so that the density matrix can be written perturbatively as $\rho_A=\rho_A^{vac}+\lambda \rho_1 + \lambda^2 \rho_2 + \dots $. In this case, the CFT state will be dual to a spacetime with metric $g_{\mu\nu}(\lambda) =g_{\mu\nu}^{(0)}+\lambda g_{\mu\nu}^{(1)}+\lambda^2g_{\mu\nu}^{(2)} + \dots$.

We recall that relative entropy vanishes up to second order in perturbations; making use of the expression (\ref{eq:deltaH}), we will check that the gravitational expression for relative entropy also vanishes up to second order for general regions $A$ bounded on a light cone. We then further make use of  (\ref{eq:deltaH}) to derive a gravitational expression dual to the first non-vanishing contribution to relative entropy, expressing it as a quadratic form in the first order metric perturbation.

\subsection{Light cone coordinates for AdS}

It will be convenient to introduce coordinates for AdS$_{d+1}$ tailored to the light cone on which the boundary of $A$ lies. Starting from standard Poincar\'e coordinates with metric
\begin{equation}
ds^2=\frac{1}{z^2}\left( dz^2-dt^2+ d \vec x^2 \right),
\end{equation}
we assume that the point $p$ whose light cone contains $\partial A$ is at $\vec{x} = z = 0$ and $t = \rho^+_0$, where $\rho^+_0$ is an arbitrary constant. On the AdS boundary, we introduce polar coordinates $(t,\rho,\Omega) = (t,\rho,\phi^1,\dots,\phi^{d-2})$ centered at $\vec{x}=0$ and define $\rho^\pm = t \pm \rho$.

The surface $\partial A$ is then described by $\rho^+ = \rho^+_0$ and some function $\rho^- = \Lambda(\phi^i)$. With these coordinates, the vector field (\ref{defZetaA}) defining the boundary modular flow takes the form
\begin{align}
\label{zeta_polar}
\xi|_{\hat{A}}&=\frac{2\pi(\rho^+_0-\rho^-)(\rho^--\Lambda(\phi^i))}{\rho^+_0-\Lambda(\phi^i)} \partial_{-}.
\end{align}
and the modular Hamiltonian (\ref{defEA}) may be written explicitly as
\begin{align}
\label{HA}
H_{A}=4\pi\iint_{\Lambda(\phi^i)}^{\rho^+_0} d\rho^-d\Omega \left( \frac{\rho^+_0-\rho^-}{2} \right)^{d-1}\left[ \frac{\rho^--\Lambda(\phi^i)}{\rho^+_0-\Lambda(\phi^i)} \right]T_{--}.
\end{align}
For the choice $\Lambda(\phi^i) = -\rho^+_0$ the region $A$ is a ball of radius $\rho^+_0$ centered at the origin on the $t=0$ slice and the expression reduces to the usual expression for a modular Hamiltonian of such a ball-shaped region.

In the bulk, we similarly define polar coordinates $(t,r,\theta,\phi^1,\dots,\phi^{d-2})$ where $(\rho,z) = r (\cos\theta, \sin \theta)$ and define $r^\pm \equiv  t \pm r$ so that the bulk light cone of the point $p$ is $r^+ = \rho^+_0$. We will see below that for pure AdS, the extremal surface $\tilde{A}$ lies on this bulk light cone on a surface that we will parameterize as $r^- = \Lambda(\theta,\phi^i)$, where $\Lambda(\theta=0, \phi^i)$ is the function that parameterized the surface $\partial A$.

The AdS$_{d+1}$ line element in these coordinates reads
\begin{align}
\label{eq:adsMetricPolarCoord}
\begin{split}
ds^2 &= \frac{1}{\sin^2 \theta} \left( -\frac{4dr^+dr^-}{(r^+ - r^-)^2} + d\theta^2 + \cos^2 \theta g_{ij}^{\Omega}d\phi^id\phi^j \right),
\end{split}
\end{align}
where $g_{ij}^{\Omega}$ is the metric on the unit $d-2$ sphere and only depends on $\phi^i$.

\subsection{HRRT surface in pure AdS}
\label{sec:hrrt_surface_in_pure_ads}
In this section, we derive an analytic expression for the extremal surface $\tilde{A}$ in pure AdS whose boundary is the region $\partial A$ on the lightcone of $p$. This will be useful in giving more explicit expressions for the relative entropy at leading order in perturbations.

We choose static gauge, parameterizing the surface using the spacetime coordinates $\theta$ and $\phi^i$ and describing its profile in the other directions by $\rho^\pm(\theta,\phi^i)$. The equations which determine its location are
\begin{align}
\label{eq:xtremalSurface}
\gamma^{ab} \frac{\partial \gamma_{ab}}{\partial r^\pm}  = - \frac 1 {\sqrt{\gamma}} \partial_a \left(  \frac{8 \sqrt{\gamma} \gamma^{ab} }{\sin^2 \theta (r^+ - r^-)^2} \partial_b r^\mp \right).
\end{align}
Let us make the ansatz that even away from the boundary the extremal surface lives on the lightcone $r^+=\rho^+_0$ and $r^-=\Lambda(\theta, \phi^i)$. The induced metric of this codimension two surface is
\begin{align}
\label{eq:induced_metric}
\gamma_{ab} = \frac{1}{\sin^2 \theta} \left( \delta^\theta_a \delta^\theta_b +  \cos^2 \theta g_{ij}^{\Omega}\delta_a^i \delta_b^i \right),
\end{align}
where $a,b \in \{\theta,\phi^1,...,\phi^{d-2}\}$ and $i,j \in\{\phi^1,...,\phi^{d-2}\}$. This metric is independent of $r^\pm$; we will see in section \ref{sec:Markov} that this is related to the Markov property of CFT subregions with boundary on a lightcone.

Since the induced metric is independent of $r^\pm$, the left hand side of the equations of motion \eqref{eq:xtremalSurface} vanishes and we can see from the right hand side that the ansatz $r^+ = \rho_0^+$ solves the equations. The remaining equation for $f(\theta, \phi^i) \equiv \rho_0^+-\Lambda(\theta,\phi^i)$ reads
\begin{align}
\label{eq:eom_extremal_surface}
0 = \partial_a \left( \frac{\sqrt{\gamma} \gamma^{ab}}{\sin^2 \theta} \partial_b \frac 1 {f(\theta,\phi^i)} \right).
\end{align}
The solution which corresponds to ball-shaped entangling surfaces is well known to be located at $\rho^2 + z^2 = \text{const}$. In order to obtain the solution for entangling surfaces of arbitrary shape (but still on a lightcone) we substitute the expression for the induced metric and separate the equation for $r^-$ into
\begin{align}
\cos^3 \theta \tan^{d-1} \theta \partial_\theta \left( \frac{1}{\cos \theta \tan^{d-1}\theta} \partial_\theta \frac 1 {f(\theta,\phi^i)} \right) = - \frac{1}{\sqrt{g^{\Omega}}} \partial_i\left( \sqrt{g^{\Omega}} (g^{\Omega})^{i j}  \partial_j \frac 1 {f(\theta,\phi^i)} \right).
\end{align}
Here, we followed our conventions and used indices $i,j$ for the angular coordinates $\phi^i$. If we write $\frac 1{f} = h(\theta) \Phi(\phi^i)$ we find that the left hand side can be solved if $\Phi(\phi^i)$ is a spherical harmonic. In $d-2$ dimensions, the eigenvalues of the Laplacian on $S^{d-2}$ are given by $n(3-d-n)$ for the $n$-th harmonic. Every level $n$ has a corresponding set of degenerate eigenfunctions $\Phi_n^l$ with $l = 1,\dots, \frac{2n+d-3}{n} \binom{n+d-4}{n-1}$ \cite{Frye:2012jj}. The left hand side reads
\begin{align}
\cos^2 \theta  h''(\theta) - \cot \theta (\cos^2 \theta + (d-2)  )  h'(\theta) + n (3 - d - n) h(\theta) = 0.
\end{align}
This differential equation can be solved in terms of hypergeometric functions,
\begin{align}
\label{eq:HypergeometricSolutionExtremalSurface}
\begin{split}
h(\theta) =& c_1 \cos ^{3-d-n} \theta \, _2F_1\left(\frac{2 - d - n}{2},\frac{3 - d - n}{2};\frac{5 - d - 2n}{2};\cos^2 \theta \right)\\
&+c_2 \cos ^n \theta \, _2F_1\left(\frac{n -1}{2},\frac{n}{2};\frac{d -1+ 2n}{2};\cos ^2 \theta \right).
\end{split}
\end{align}
To fix the constants in \eqref{eq:HypergeometricSolutionExtremalSurface} it helps to use intuition from the solutions in the case where the boundary of a subregion is located on a null-plane instead of a lightcone (see appendix B). In that case it is clear that effects from perturbations away from a constant entangling surface on the extremal surface die off as $z \to \infty$. Under a transformation which maps the Rindler result to a ball-shaped region, the distant part of the extremal surface corresponds to $\theta = \pi/2$. Consequently, we require that $h_n(\pi/2) \to 0$ for $n \geq 1$ and $h_n(\pi/2)=1$ for $n=0$. At the same time, for $\theta \to 0$ we need that $h_n(\theta)$ is constant and different from zero. These constraints are easily solved with $c_1 = 0, c_2=1$. Introducing a normalization factor to ensure that $h_n(0) = 1$, we are left with
\begin{align}
\label{eq:extremalSurface}
h_n(\theta)= \cos^n \theta \frac{\Gamma(\frac {d+n} 2) \Gamma(\frac {d -1 + n} 2) }{\Gamma(\frac {d-1} 2 + n)\Gamma(\frac {d} 2)} \, _2F_1\left(\frac{n-1}{2},\frac{n}{2};\frac{d-1+2n}{2};\cos^2 \theta \right) \; .
\end{align}

In conclusion this shows that extremal surfaces in the bulk are located at $r^+ = \rho_0^+$ and $r^- = \Lambda(\theta, \phi^i)$ with
\begin{align}
\Lambda(\theta, \phi^i) = \rho_0^+ - \frac{1}{C_0+\sum_{n=1}^{\infty}\sum_{l} C_{n,l} h_n(\theta)\Phi_n^l(\phi^i)}.
\end{align}
Here, $n$ runs over spherical harmonics in $d-2$ dimensions and $l$ over their respective degeneracy. They intersect the boundary at
\begin{align}
\Lambda(\phi^i) = \rho_0^+ - \frac{1}{C_0+\sum_{n=1}^{\infty}\sum_{l} C_{n,l} \Phi_n^l(\phi^i)}.
\end{align}
Thus, the constants $C_{n,l}$ are determined in terms of the function parameterizing the boundary surface by performing the spherical harmonic expansion
\be
\frac{1}{\rho_0^+ - \Lambda(\phi^i)} = C_0+\sum_{n=1}^{\infty}\sum_{l} C_{n,l} \Phi_n^l(\phi^i) \; .
\ee

As a simple example, one choice of surface involving only the $n=1$ harmonics for the AdS$_4$ case takes the form
\begin{align}
\label{eq:boostedBall}
\rho^+(\phi) = \rho_0^+, && \rho^-(\phi) = \rho_0^+ - \frac{2 \rho_0^+ \sqrt{1 - \beta ^2}}{1 + \beta \cos \phi},
\end{align}
and correspond to ball-shaped regions in a reference frame boosted relative to the original one by velocity $\beta$ in the $x$-direction.

\subsection{The bulk vector field}

Our next step is to provide an explicit expression for the vector field on the extremal surface which obeys equations \eqref{A3} -- \eqref{A2}, such that the quantity $H_\xi$ is dual to relative entropy.

Using (\ref{zeta_polar}), the explicit form of equation \eqref{A3} is
\begin{align}
\label{BCs3}
\xi|_{\hat{A}}&=\frac{2\pi(\rho^+_0-\rho^-)(\rho^--\Lambda(0,\phi^i))}{\rho^+_0-\Lambda(0,\phi^i)} \partial_{-}.
\end{align}
Equation \eqref{A1} requires knowledge of the unit binormal
\begin{equation}
n^{\mu\nu}=n_2^{\mu}n_1^{\nu}-n_1^{\mu}n_2^{\nu},
\end{equation}
but thanks to the knowledge about the expression for the extremal surface which we found in the preceding section it is possible to calculate it explicitly. Here, $n_{1,2}$ denote two orthogonal normal vectors to the RT surface. The calculation is delegated to appendix \ref{app:binormal}. The non-zero components of the unit binormal read
\begin{align}
n^{+-}=g^{+-}, && n^{a -}=-\partial^{a}\Lambda(\theta,\phi^i),
\end{align}
where $a$ again runs over coordinates $(\theta, \phi^i)$. One possible choice of a vector field satisfying the boundary conditions given by equations \eqref{A1} is:\footnote{Upon expanding the sums in equation \eqref{eq:vectorfield} it looks like the $\phi^i$ components of the vector field diverge as $\theta \to \frac \pi 2$ and for $d>3$ as $\phi^i \to 0,\pi$ due to the metric on the $S^{d-2}$ sphere. However, these divergences can be shown to be mere coordinate singularities: From equation \eqref{eq:extremalSurface} we see that $\partial_i\Lambda \sim \cos \theta$. Hence the $\phi_i$ components of the vector field go only as $\cos^{-1} \theta$. This happens as a consequence of the coordinate singularity at $\theta = \pi/2$ in polar coordinates which can be removed by going into Poincar\'e coordinates $(t,z,\vec{x})$. Similar arguments also hold for singularities due to the $S^{d-2}$ metric. Coordinate independent quantities like the norm of the spatial part of the vector field remain finite as can be seen from inspecting the metric. }
\begin{align}
\label{eq:vectorfield}
\begin{split}
\xi=&\frac{2\pi(\rho^+_0-r^+)(r^+-\Lambda(\theta,\phi^i))}{\rho^+_0-\Lambda(\theta,\phi^i)}\partial_{+} +\frac{2\pi(\rho^+_0-r^-)(r^--\Lambda(\theta,\phi^i))}{\rho^+_0-\Lambda(\theta,\phi^i)}\partial_{-}
\\ & \qquad + \frac{4\pi(\rho^+_0-r^+)}{\sin^2 \theta}\partial^a\left( \frac{1}{\rho^+_0-\Lambda(\theta,\phi^i)} \right)\partial_a.
\end{split}
\end{align}
Here, the $\partial_-$ and $\partial_+$ components are chosen to match with the expression for the Killing vector $\xi$ in the case when $\Lambda$ is constant. On the light cone, only the $\partial_-$ component (along the lightcone) is nonzero, and this has the same qualitative behavior as the vector $\zeta$ on the boundary lightcone. It is immediately clear that conditions \eqref{A3} and \eqref{A2} are satisfied. It is also straightforward to verify the condition involving the unit binormal using the fact that for a torsion free connection we have $\nabla_{\mu}\xi_{\nu}-\nabla_{\nu}\xi_{\mu} =\partial_{\mu}\xi_{\nu}-\partial_{\nu}\xi_{\mu}$.

Calculating the Lie derivative of the metric with respect to this vector field gives zero on the light cone $r^+ = \rho_0^+$ but not away from the light cone. This is in contrast to the case of a ball-shaped region, where the Lie derivative vanished everywhere inside the entanglement wedge.

\subsection{Perturbative formulae for $\Delta H_\xi$}

To write an explicit perturbative expression for $\Delta H_\xi$, we begin with the on-shell result
\be
\delta H_{\xi}=\int_{\Sigma}\omega\left( g(\lambda),\frac{d}{d\lambda}g,\mathcal{L}_{\xi}g(\lambda) \right) \; .
\ee
Here, the symplectic $d$-form $\omega$ is explicitly given by
\begin{align}
\omega\left(g(\lambda),\frac{d}{d\lambda}g,\mathcal{L}_{\xi}g(\lambda)\right) &=\frac{1}{16\pi G_N}\hat \epsilon_{\mu}P^{\mu\nu\alpha\beta\sigma\rho}\left( \mathcal{L}_{\xi}g_{\nu\alpha}\nabla_{\beta}\frac{d}{d\lambda} g_{\sigma \rho}-\frac{d}{d\lambda}g_{\nu\alpha}\nabla_{\beta}\mathcal{L}_{\xi}g_{\sigma\rho} \right),
\end{align}
where
\begin{align}
\begin{split}
P^{\mu\nu\alpha\beta\sigma\rho}&=g^{\mu\sigma}g^{\nu\rho}g^{\alpha\beta}-\frac{1}{2}g^{\mu\beta}g^{\nu\sigma}g^{\rho\alpha}-\frac{1}{2}g^{\mu\nu}g^{\alpha\beta}g^{\sigma\rho}-\frac{1}{2}g^{\nu\alpha}g^{\mu\sigma}g^{\beta\rho} +\frac{1}{2}g^{\nu\alpha}g^{\mu\beta}g^{\sigma\rho}.
\end{split}
\end{align}
Since both $P^{\mu\nu\alpha\beta\sigma\rho}$ and $\hat \epsilon_{\mu}$ depend on the metric they will have a series expansions in $\lambda$ when we express the metric as a series. Also in this case we will use sub- or superscripts in parenthesis to indicate the order of the term in $\lambda$. Here and in the following we will use $\nabla_\mu$ to denote covariant derivatives with respect to $g_{\mu\nu}(0)=g^{(0)}_{\mu\nu}$.

It will be convenient for us to choose a gauge for the metric perturbations such that the extremal surface stays at the same coordinate location for any variation of the metric. It was shown in \cite{Hollands:2012sf} that this is always possible. In this case, we have at first order
\be
\label{REfirst}
\Delta H_{\xi}^{(1)}=\int_{\Sigma}\omega\left( g(\lambda),\frac{d}{d\lambda}g,\mathcal{L}_{\xi} g(\lambda) \right) \; .
\ee
We will see in the next section that this vanishes, in accord with the general vanishing of relative entropy at first order (also known as the first law of entanglement).

At second order, we have
\begin{align}
\label{REsecond}
\frac{d^2}{d\lambda^2}S(g(\lambda)||g_0)|_{\lambda=0}=\int_{\Sigma}\frac{d}{d\lambda}\omega\left(g(\lambda),\frac{d}{d\lambda}g,\mathcal{L}_{\xi}g(\lambda)\right)\biggr\vert_{\lambda=0}.
\end{align}
We will calculate this more explicitly in section (\ref{sec:secondOrder}).

\subsubsection{Vanishing of the first order expression}
\label{sec:firstOrder}
In this section, we demonstrate that our gravitational expression for the relative entropy vanishes for first order perturbations as required. Expanding the first order expression (\ref{REfirst}) for $\omega$ yields
\begin{equation}
\int_{\Sigma}\omega\left(g(\lambda),\frac{d}{d\lambda}g,\mathcal{L}_{\xi}g(\lambda)\right)\biggr\vert_{\lambda=0} =-\frac{1}{32 \pi G_N}\int_{\Sigma}\epsilon^{(0)}_+\left(g_{(0)}^{+-}\right)^2g^{(1)}_{--} g^{ab}_{(0)}\partial_+\mathcal{L}_{\xi}g^{(0)}_{ab},
\end{equation}
where repeated lower case letters $a,b$ imply summation over angular coordinates $(\theta,\phi^i)$. Using the definition of the Lie derivative
\be
g_{(0)}^{ab}\partial_+\mathcal{L}_{\xi}g^{(0)}_{ab} = 2 g_{(0)}^{ab} \partial_+\nabla_a \xi_b,
\ee
and the fact that since $g^{ab}$ is independent of $r^\pm$ all Christoffel symbols of the form $\Gamma_{ab}^\pm$ vanish at leading order, the problem reduces to a problem of only the angular coordinates. We obtain
\be
g_{(0)}^{ab}\partial_+\mathcal{L}_{\xi}g^{(0)}_{ab} = 2 \nabla_a \partial_+\xi^a = \frac{2}{\sqrt{\gamma}} \partial_a \left( \sqrt{\gamma} \partial_+ \xi^a \right).
\ee
Substituting the general form of $\xi^a$ from equation \eqref{eq:vectorfield} and using that $g^{(0)}_{ab} = \gamma^{(0)}_{ab}$ we end up with
\begin{align}
\gamma_{(0)}^{ab}\partial_+\mathcal{L}_{\xi}\gamma^{(0)}_{ab} = -8\pi \frac{1}{\sqrt{\gamma}} \partial_a \left(  \frac{\sqrt{\gamma} \gamma_{(0)}^{ab}}{\sin^2 \theta} \partial_b \frac{1}{\rho_0^+-\Lambda} \right).
\end{align}
$g_{\mu\nu}^{(0)}$ and $g_{\mu\nu}^{(1)}$ are the bulk metric and its perturbation and $\gamma_{ab}^{(0)}$, $\gamma_{ab}^{(1)}$ are the induced metric and the induced metric perturbation, respectively.
This expression is proportional to the equation for an extremal surface, equation \eqref{eq:eom_extremal_surface}, and therefore vanishes.

If we drop the assumption that the Einstein equations are satisfied, one can show that the first law of entanglement entropy implies that the Einstein equations hold at first order around pure AdS. This was done in \cite{Faulkner:2013ica} where only ball-shaped CFT subregions were considered. Utilizing more general subregions bounded by a lightcone does not yield new (in-)equalities at first order. 

\subsubsection{Relative entropy at second order}
\label{sec:secondOrder}
We will now provide a more explicit expression for the leading perturbative contribution to relative entropy, which appears at second order in the perturbations. Starting from (\ref{REsecond}) and using our explicit expression for $\omega$, we obtain four potentially contributing terms,
\begin{align}
\begin{split}
 \frac{d^2}{d\lambda^2}S(g(\lambda)||g_0)|_{\lambda=0}=&\frac{1}{16\pi G_N}\int_{\Sigma}\epsilon_+^{(1)}P_{(0)}^{+\nu\alpha\beta\sigma\rho}\left( \mathcal{L}_{\xi}g^{(0)}_{\nu\alpha}\nabla_{\beta}g^{(1)}_{\sigma\rho} -g^{(1)}_{\nu\alpha}\nabla_{\beta}\mathcal{L}_{\xi}g^{(0)}_{\sigma\rho} \right) \\
 & +\frac{1}{16\pi G_N}\int_{\Sigma} \epsilon^{(0)}P^{+\nu\alpha\beta\sigma\rho}_{(1)} \left( \mathcal{L}_{\xi}g^{(0)}_{\nu\alpha}\nabla_{\beta}g^{(1)}_{\sigma\rho} -g^{(1)}_{\nu\alpha}\nabla_{\beta}\mathcal{L}_{\xi}g^{(0)}_{\sigma\rho} \right) \\
 & +\frac{1}{16\pi G_N} \int_{\Sigma} \epsilon^{(0)}P^{+\nu\alpha\beta\sigma\rho}_{(0)} \left( \mathcal{L}_{\xi}g^{(0)}_{\nu\alpha}\nabla_{\beta}g^{(2)}_{\sigma\rho} -g^{(2)}_{\nu\alpha}\nabla_{\beta}\mathcal{L}_{\xi}g^{(0)}_{\sigma\rho} \right)\\
 & +\frac{1}{16\pi G_N} \int_{\Sigma} \epsilon^{(0)}P^{+\nu\alpha\beta\sigma\rho}_{(0)} \left( \mathcal{L}_{\xi}g^{(1)}_{\nu\alpha}\nabla_{\beta}g^{(1)}_{\sigma\rho} -g^{(1)}_{\nu\alpha}\nabla_{\beta}\mathcal{L}_{\xi}g^{(1)}_{\sigma\rho} \right) .
 \end{split}
 \end{align}
 The first and third terms vanish because of our first order results of section \ref{sec:firstOrder}. The last term is reminiscent of the standard canonical energy associated with the interior of the entanglement wedge, except that $\xi$ is no longer a Killing vector. The non zero contributions take the form
\begin{align}
\begin{split}
\delta^{(2)}H_\xi =&\int_{\Sigma}\omega\left(g(\lambda),\frac{d}{d\lambda}g,\mathcal{L}_{\xi}\frac{d}{d\lambda}g\right)\biggr\vert_{\lambda=0}\\
&-\frac{1}{16\pi G_N}\int_{\Sigma}\epsilon^{(0)}_+\left(g^{+-}_{(0)}\right)^2\left[ g^{(1)}_{-c}g_{(0)}^{ca}g_{(0)}^{db}g^{(1)}_{-d}-g^{(1)}_{--}g^{ab}_{(1)}\right]\partial_+\mathcal{L}_{\xi}g^{(0)}_{ab}
\end{split}
\label{final}
\end{align}
Here, $a,b,c,d$ run over angular coordinates, $\mu,\nu$ run over all coordinates. Note that although we are calculating relative entropy at second order, the expression only depends on first order metric perturbations. Due to the fact that $\xi$ is no longer a Killing vector field, we appear to have a contribution in addition to the first term which appears for the case of ball-shaped regions.

However, we have not yet imposed the Hollands-Wald gauge condition on the first order metric perturbations, for which the coordinate location of the extremal surface is the same as in the case of pure AdS. We have additional gauge freedom on top of this, and it may be that for a suitable gauge choice, the final term in the expression above can be eliminated. We have checked that this is the case for a planar black hole in AdS$_4$. We discuss this, as well as the procedure of choosing the Hollands-Wald gauge condition in more detail in appendix D.

\section{Holographic proof of the Markov property of the vacuum state}
\label{sec:Markov}
In \cite{Casini:2017roe} it was pointed out that the vacuum states of subregions of a CFT bounded by curves $\rho^- = \Lambda_A$ and $\rho^- = \Lambda_B$ on the lightcone $\rho^+ = \rho^+_0$ saturate strong subadditivity, i.e.
\begin{align}
S_{A} + S_{B} - S_{A \cap B} - S_{A \cup B} = 0.
\end{align}
This is also known as the Markov property. Moreover, even for CFTs deformed by relevant perturbations, the reduced density matrices for regions $A$ and $B$ describe Markov states if $A$ and $B$ have their boundary on a null-plane. In its most general form the proof used that the modular Hamiltonians for such regions obey
\begin{align}
H_{A} + H_{B} - H_{A \cap B} - H_{A \cup B} = 0,
\end{align}
which can be proven using methods of algebraic QFT. In this section we will give a holographic proof of the Markov property which uses the Ryu-Takayanagi proposal for entanglement entropy. We will start with the proof for a subregion of a deformed CFT with boundary on a null-plane and after that also show the property for subregions of CFTs with boundary on a lightcone.

\subsection{The Markov property for states on the null-plane}
The vacuum state of a deformed CFT is dual to a geometry of the form
\begin{align}
\begin{split}
ds^2 &= f(z)dz^2 + g(z)(-2dx^+dx^- + dx_\perp^\mu dx_{\perp \mu}).
\end{split}
\end{align}
An undeformed CFT corresponds to the special case $f(z) = g(z) = \frac 1 {z^2}$. The entanglement entropy of a subregion $A$ can then be calculated using the RT prescription, following the same steps as in section \ref{sec:HRRT_on_the_null_plane}. We assume that the boundary $\partial A$ is described by $x^- = \text{const}$ and $x^+ = x^+(\vec x_\perp)$. To describe the corresponding extremal surface we go to static gauge, where $z$ and $x_\perp$ are our coordinates and $x^{\pm}(z,x_\perp)$ is the embedding. 
The ansatz $x^- = \rm const$ and $x^+ = x^+(\vec x_\perp,z)$ simplifies the equation to
\begin{align}
0 = \partial_a (\sqrt{\gamma} \gamma^{ab} \partial_b x^+ g_{+ -}).
\end{align}
The relevant solution to this equation in the case of pure AdS is discussed in appendix \ref{sec:HRRT_on_the_null_plane} and is given by
\begin{align}
x^+(z,x^i_\perp) = \frac{2^{\frac{2-d}2}k^{d/2}}{\Gamma(d/2)} \int d^{d-2}k a_{k^i} z^{d/2}  K_{d/2} (z k) e^{i k^i x^i}.
\end{align}
Here, $K_{d/2}$ is the modified Bessel function of the second kind and the coefficients $a_{k^i}$ are given in terms of the entangling surface $x^+(0,x^i_\perp)$ as
\begin{align}
a_{k} = \int \frac{d^{d-2}x_\perp}{(2 \pi)^{d-2}} e^{-i k \cdot x_\perp} x^+(0,x^i_\perp).
\end{align}
More generally, the induced metric on the extremal surface in the bulk is
\begin{align}
ds^2 = f(z)dz^2 + g(z)(dx_\perp^\mu dx_{\perp \mu})
\end{align}
and independent of the embedding $x^+(\vec x_\perp,z)$. Thus, it is clear that the areas of all extremal surfaces ending on $x^-=\text{const}$ are the same, potentially up to terms which depend on how the area of the extremal surface is regularized as we approach the boundary. The standard prescription given by cutting off $z$ at some distance $\epsilon$ away from the boundary gives a universal cutoff term for all such extremal surfaces and therefore the entanglement entropies for all regions with boundary on $x^-$ are identical and strong subadditivity is saturated. Our argument is an explicit version of very similar arguments which have been used to show the saturation of the Quantum Null Energy condition \cite{Koeller:2017njr}.\footnote{We thank Adam Levine for pointing this out to us.}

\subsection{The Markov property for states on the lightcone}
If we consider an arbitrary region on the lightcone we expect the Markov property to hold for undeformed CFTs, since the lightcone is conformally equivalent to the null-plane. The solution for an extremal surface in pure AdS ending on a lightcone at the boundary was already discussed in section \ref{sec:hrrt_surface_in_pure_ads}. Consider the case where we have two different entangling surfaces given by $\rho^- = \Lambda_A(\phi_i)$ and $\rho^- = \Lambda_B(\phi_i)$. We have seen before that the metric on the extremal surface is in fact $r^-$ independent. However, again the dependence on the entangling surface can enter through regularization of the integral and would show up in the cutoff-dependent term.

In the coordinates of our choice $\theta, \phi^i$ the divergent term in the area comes from the integral over $\theta$. Following the standard way of regulating the surface integral we introduce a cutoff $z = \epsilon$, which translates into cutting off the integral at $\theta = \frac \epsilon r \approx \frac \epsilon \rho$. From this is follows that if we choose the canonical way of regulating the entropy, the $\theta$ integral runs from $\frac{2 \epsilon}{(\rho^+_0 - \Lambda)} \equiv \theta_-$ to $\pi/2$.

The entropy which is proportional to the area term can now be calculated using the explicit form of the induced metric, equation \eqref{eq:induced_metric}, and is given by
\begin{align}
\label{eq:MarkovConeRelEntropy}
\begin{split}\int \sqrt{\gamma} &= \int d\Omega \int_{\theta_-}^{\pi/2}d\theta \frac{\cos^{d-2} \theta}{\sin^{d-1} \theta}.
\end{split}
\end{align}
The only way the shape of the entangling surface appears is through the cutoff, i.e.~the surface area can be expanded as
\begin{align}
A = \sum_{\alpha = d-2}^0 c_n \left( \frac{\rho^+_0 - \Lambda(\phi^i)}{2 \epsilon} \right)^{\alpha},
\end{align}
where the coefficients $c_n$ are the same for all entangling surfaces.
In the light of equation \eqref{eq:MarkovConeRelEntropy} saturation of strong subadditivity for two regions on a lightcone defined by $\Lambda_A$ and $\Lambda_B$ is guaranteed if
\begin{align}
\begin{split}
\int d\Omega & \left( (\rho^+_0 -\Lambda_A(\phi^i))^{\alpha} + (\rho^+_0 -\Lambda_B(\phi^i))^{\alpha} \right.\\ &  \left.- \max(\rho^+_0 -\Lambda_A(\phi^i), \rho^+_0 -\Lambda_B(\phi^i))^{\alpha} - \min(\rho^+_0 -\Lambda_A(\phi^i), \rho^+_0 -\Lambda_B(\phi^i))^{\alpha}\right) = 0,
\end{split}
\end{align}
which is trivially pointwise true. This again shows that strong subadditivity is saturated, or in other words, reduced density matrices for regions on the lightcone describe Markovian states. For more details on the form of the coefficients $c_n$ in the expansion, see \cite{Casini:2018kzx}.

The authors of \cite{Casini:2017roe} also speculated about the possibility of introducing a cutoff to regulate the area of extremal surfaces such that the area of the extremal surfaces of subregions on the lightcone are all exactly equal. The previous discussion explicitly shows that choosing to introduce a cutoff $\theta = \epsilon$ instead of $z = \epsilon$ realizes such a regularization procedure in which all entanglement entropies for regions on the lightcone are in fact the same.

\section{Discussion}

The results of this paper imply that for any classical asymptotically AdS spacetime arising in a consistent theory of quantum gravity, the energy $\Delta H_\xi$ must be positive and must not decrease as we increase the size of region $A$. It would be interesting to understand if it is possible to prove this result directly in general relativity, by requiring that the matter stress-energy tensor satisfy some standard energy condition.

It may be useful to point out that there is a differential quantity whose positivity implies all the other positivity and monotonicity results considered here. If we consider a deformation of the region $A$  by an infinitesimal amount $\epsilon v(\Omega)$, where $v$ is some vector field on $\partial A$ pointing along the lightcone away from $p$, the change in relative entropy to first order must take the form
\be
\delta S(\rho_A || \rho_A^{vac}) = \epsilon \int \d \Omega v(\Omega) S_A(\Omega)
\ee
The monotonicity property implies that the quantity $S_A(\Omega)$ must be positive for all $A$ and all $\Omega$.\footnote{A special case of this positivity result was utilized in the proof of the averaged null energy condition in \cite{Faulkner:2016mzt}.} It would be interesting to make use of our results to come up with a more explicit expression for the gravitational analogue of the quantity $S_A(\Omega)$. One approach to providing a GR proof of the subsystem energy theorems would be to prove positivity of this.

The Markov property discussed in section 5 suggests that it should be interesting to consider (for general states) the gravitational dual of the combination $S(A)+ S(B) - S(A \cup B) - S(A \cap B)$ of entanglement entropies for regions $A$ and $B$ on a lightcone. Since strong subadditivity is saturated for the vacuum state, this gravitational quantity will vanish for pure AdS, but must be positive for any nearby physical asymptotically AdS spacetime according to strong subadditivity. Thus, strong subadditivity for these regions on a light cone will lead to a constraint on gravitational physics that appears even when considering small perturbations away from AdS. For two-dimensional CFTs, this quantity was already considered previously in \cite{Lashkari:2014kda,Banerjee:2014oaa}; the analysis there suggests that this gravitational constraint takes the form of a spatially integrated null-energy condition. See \cite{VanRaamsdonk:2016exw} for some additional discussion of gravitational constraints from strong subadditivity.

\section*{Acknowledgements}
We would like to thank Alex May for helpful discussions. This work was supported in part by the Natural Sciences and Engineering Research Council of Canada and by the Simons Foundation. DN is supported by a UBC Four
Year Doctoral Fellowship.
\appendix

\section{Equivalence of $H_{\xi}$ on the boundary and the modular Hamiltonian}
\label{app:modularHamiltonian}
In this appendix we will show that $H_{\xi}$ reduces to the modular Hamiltonian on the boundary, even in the case of a deformed entangling surface. We take the infinitesimal difference between pure AdS and another spacetime that satisfies the linearized Einstein's equations around pure AdS, i.e. we want to calculate $\delta Q_{\xi}-\xi \cdot \theta$ on a constant $z$ slice near the boundary. We can find in the appendix of \cite{Lashkari:2016idm} that
\begin{equation}
\delta Q_{\xi}-\xi \cdot \theta = \frac{1}{16\pi G_N} \hat{\epsilon}_{ab}\left[ \delta g^{ac}\nabla_c \xi^{b}-\frac{1}{2}\delta g^c_c \nabla^{a}\xi^b +\xi^c\nabla^b\delta g^a_c - \xi^b\nabla_c\delta g^{ca} +\xi^b\nabla^a \delta g^c_c \right].
\end{equation}
The next step is to expand the sum over $a$ and $b$. As we approach the boundary we consider volume elements on on constant $z$ slices and thus the term involving the volume element $\hat{\epsilon}_{\mu \nu}$ vanishes. In Fefferman-Graham gauge ($\delta g_{zc}=0$) we find
\begin{equation}
\label{eq:nearBdry}
\begin{split}
\delta Q_{\xi}-\xi \cdot \theta&=\frac{1}{16 \pi G_N}\hat{\epsilon}_{\mu z}\left[ \frac{1}{2}\delta g^{\nu}_{\nu}\nabla^z\xi^{\mu}-\xi^{\mu}\nabla^z\delta g^{\nu}_{\nu}-\xi^c\nabla^{\mu}\delta g^z_c+\xi^{\mu}\nabla_c \delta g^{cz} \right]\\
&+\frac{1}{16\pi G_N} \hat{\epsilon}_{\mu z}\left[ \delta g^{\mu \nu}\nabla_{\nu} \xi^{z}-\frac{1}{2}\delta g^{\nu}_{\nu} \nabla^{\mu}\xi^z +\xi^{\nu}\nabla^z\delta g^{\mu}_{\nu} - \xi^z\nabla_{\nu}\delta g^{\nu \mu} +\xi^z\nabla^{\mu} \delta g^{\nu}_{\nu} \right].
\end{split}
\end{equation}
Now all we need to do is find the leading order behaviour near $z=0$. To this effect we assume that the vector fields have a asymptotic expansion near the conformal boundary given in equation \eqref{eq:behaviorVectorField}.

We also take $\delta g_{ab}=z^{d-2}\Gamma^{(d)}_{ab}+z^{d-1}\Gamma^{(d+1)}_{ab}+...$. The leading order terms of equation \eqref{eq:nearBdry} are
\begin{equation}
\frac{d}{16\pi G_N} \eta^{\mu \lambda}
\hat{\epsilon}_{\mu z} \Gamma^{(d)}_{\lambda \nu}\xi^{\nu}z^{d+1}+...=\mathcal{O}(1),
\end{equation}
where we use the fact that for a CFT traceless stress-energy tensor implies that $\eta^{\nu \rho} g^{(d)}_{\nu \rho}=0$ and $\hat{\epsilon}_{\mu z}=\mathcal{O}\left(z^{-(d+1)}\right)$. Finally, employing the relation between the metric perturbation in FG coordinates and the stress-energy tensor,
\begin{align}
\Delta\langle T_{\mu\nu} \rangle = \frac{d}{16\pi G_N}\left. \Gamma^{(d)}_{\mu \nu}\right|_{z=0}
\end{align}
and the definition of $\epsilon$ given in section \ref{sec:background} we arrive at
\begin{equation}
\delta Q_{\xi}-\xi \cdot \theta = {\epsilon}^{\rho}\left< T_{\rho \sigma} \right>\xi^{\sigma}+\mathcal{O}(z).
\end{equation}

\section{The HRRT surface ending on the null-plane}
\label{sec:HRRT_on_the_null_plane}
In order to derive the HRRT surface which ends on a curve located on a boundary null-plane, we split the coordinates into $x^\pm = t \pm x$ (here $x$ is the spatial direction parallel to the null-plane), boundary directions $x^i_\perp$ orthogonal to the null-plane, and the bulk coordinate $z$. The metric on the Poincar\'e patch in these coordinates is
\begin{align}
ds^2 = \frac 1 {z^2}(dz^2 - 2 dx^+ dx^- + dx_\perp^i dx_{\perp i}).
\end{align}
We choose static gauge for the coordinates on our extremal surface, such that $x^\pm = x^\pm(z,x^i_\perp)$. The entangling surface on the boundary is then given by $x^\pm = x^\pm(0,x^i_\perp)$. The equations which determine the embeddings $x^\pm(z,x^i_\perp)$ are given by
\begin{align}
\label{eq:xtremalSurfacePlane}
\gamma^{ab} \frac{\partial \gamma_{ab}}{\partial x^\pm}  = - \frac 1 {\sqrt{\gamma}} \partial_a \left(  2 \sqrt{\gamma} \gamma^{ab} g_{+-} \partial_b x^\mp \right),
\end{align}
where the induced metric is denoted by $\gamma_{ab}$. Having the extremal surface ending on a boundary null-plane means that either $x^+$ or $x^-$ are constant. Without loss of generality, we choose $x^- = x^-_0 = \text{const}$. This reduces the two equations \eqref{eq:xtremalSurfacePlane} to a single equation for $x^+(z,x^i_\perp)$. Making the ansatz $x^+(z,x^i_\perp) = h_{k}(z) g_{k}(x^i_\perp)$ we can separate the equation into
\begin{align}
z^{d-1} \partial_z ( z^{1-d} \partial_z h_k(z)) = -\Delta_\perp g_{k}(x_\perp^i).
\end{align}
The general solutions for the functions $h_k(z)$ and $g_k(x^i_\perp)$ are given by
\begin{align}
g_{k}(x^\perp) &= a_{k^i} e^{i k^i x_\perp^i},\\
h_k(z) &= c_k z^{d/2} I_{d/2} (z k) + d_k z^{d/2} K_{d/2} (z k),
\end{align}
where $k = | k^i |$ and $x_\perp^i k^i$ denotes the Euclidean inner product between the vectors $k^i$ and $x^i$. $I_\nu$ and $K_\nu$ denote the modified Bessel functions of first and second kind, respectively. We also define $h_0 = \lim_{z \to 0} h_k(z)$. It turns out that we do not want the full solution for $h_k$. Intuitively, it is clear that the effect of deformations of the entangling surface on the boundary should die off as $z \to \infty$. At the same time we also require that the shape of the extremal surface is uniquely determined by boundary conditions. The asymptotic behavior of $h_k$ as $z \to \infty$ and $z\to 0$ is
\begin{align}
\lim_{z \to \infty} h_k(z) &= c_k \sqrt{\frac{1}{2 \pi k}} e^{kz} + d_k \sqrt{\frac{\pi}{2 k}} e^{-kz},\\
\lim_{z \to 0} h_k(z) &= d_k 2^{\frac {d-2}{2}} \Gamma\left(\frac d 2 \right) k^{-d/2}.
\end{align}
We can only fulfill above requirements if we set $c_k$ = 0. Hence any extremal surface ending on the null-plane $x^-=x^-_0$ is given by
\begin{align}
x^+(z,x^i_\perp) = \frac{2^{\frac{2-d}2}k^{d/2}}{\Gamma(d/2)} \int d^{d-2}k a_{\vec k} z^{d/2}  K_{d/2} (z k) e^{i k^i x^i}.
\end{align}
The normalization is chosen such that
\begin{align}
\lim_{z \to 0} x^+(z,x^i_\perp) = \int d^{d-2}k a_{k^i} e^{i k^i x_\perp^i}
\end{align}
determines $a_{k}$ in terms of the entangling surface $x^+(0,x^\perp)$.

\section{Calculation of the binormal}
\label{app:binormal}
The binormal $n^{\mu\nu}$ is defined as
\begin{align}
n^{\mu\nu}=n_2^\mu n_1^\nu-n_2^\nu n_1^\mu,
\end{align}
where $n_1$ and $n_2$ are orthogonal $\pm1$ normalized normal vectors to the extremal surface.
To calculate them start by calculating the $d-1$ tangent vectors to the surface which will be labeled by $n$ as $t_n=t_n^{\mu}\partial_{\mu}$, $n \in \{ 1,2,...,d-1 \}$. They satisfy $t_n^{\mu}\partial_{\mu}(r^+-\rho^+_0)=0$ and $t_n^{\mu}\partial_{\mu}(r^--\Lambda(\theta,\phi^i))=0$. A convenient set of tangent vectors is given by
\begin{align}
t_1 &= \sqrt{g^{\theta\theta}}\left((\partial_\theta\Lambda)\partial_- + \partial_{\theta}\right), \\
t_2 &= \sqrt{g^{\phi^1\phi^1}}\left((\partial_{\phi^1}\Lambda)\partial_-+\partial_{{\phi^1}}\right),\\
t_3 &= \sqrt{g^{\phi^2\phi^2}}\left((\partial_{\phi^2}\Lambda)\partial_-+\partial_{{\phi^2}}\right),\\
t_4 &= \dots
\end{align}
and so on for all $\phi^i$. It is easy to see that these vectors form an orthonormal basis on the Ryu-Takayanagi surface. Requiring that $n_1$ and $n_2$ are orthogonal to all tangent vectors, $g_{\mu\nu}n^{\mu}_{1,2}t_a^{\nu}=0$. This requirement is fulfilled by choosing
\begin{align}
n_{1,2}^+=g^{+-}, && n^{a}_{1,2}=- \partial^a\Lambda,
\end{align}
where $a$ stands again for all angular components. The condition that $n_1$ and $n_2$ be orthogonal and normalized to $+1$ and $-1$, respectively, is obeyed provided we choose
\begin{align}
n_1^- =\frac 1 2 \left(1- \partial^a\Lambda \partial_a\Lambda \right), &&
n_2^- =-\frac 1 2 \left(1+  \partial^a\Lambda \partial_a\Lambda \right).
\end{align}
One can check that the only non-zero components of the binormal are given by:
\begin{align}
n^{+-} =g^{+-}, &&  n^{a -} =- \partial^{a}\Lambda.
\end{align}

\section{Hollands-Wald gauge condition}

In this appendix, we argue that for the example of a planar black hole in AdS$_4$, considered as a perturbation of pure AdS,  we can choose a gauge where $g^{(1)}_{-a}|_{\Sigma} = 0 = g^{(1)}_{--}|_{\Sigma}$ which at the same time is compatible with Hollands-Wald gauge. In this case, the final term in our second order expression (\ref{final}) for the relative entropy vanishes.

Hollands-Wald gauge is determined by requiring that the extremal surface in the deformed spacetime sits at the same coordinate location than the extremal surface in the undeformed spacetime. In particular this means that
\begin{align}
r^- = \Lambda(\theta,\phi), && r^+ = \rho_0^+.
\end{align}
The requirement that also after a perturbation of the metric the extremal surface $\tilde A$ sits at its old coordinate location translates into
\begin{align}
0 ={} &\partial_- \left( \gamma_{(0)}^{ab} \gamma^{(1)}_{ab} \right) -  \partial_c(\sqrt{\gamma^{(0)}} \gamma_{(0)}^{ca} \partial_a x^\mu g^{(1)}_{-\mu})\bigg\vert_{\tilde A}, \\
\begin{split}
\label{eq:HWGaugeCondition}
0 ={} & - \frac 1 2 \sqrt{\gamma^{(0)}} \gamma_{(0)}^{ab} \partial_a r^- g^{(0)}_{+-} \partial_b(\gamma_{(0)}^{cd} \gamma^{(1)}_{cd}) + \partial_c (\sqrt{\gamma^{(0)}}  \partial_d r^- g^{(0)}_{+-} \gamma_{(0)}^{ca} \gamma^{(1)}_{ab} \gamma_{(0)}^{bd})  \\
&  - \partial_b(\sqrt{\gamma^{(0)}} \gamma_{(0)}^{ab} \partial_a r^- g^{(1)}_{+-}) - \partial_b(\sqrt{\gamma^{(0)}} \gamma_{(0)}^{ab} g^{(1)}_{+a}) + \frac 1 2 \sqrt{\gamma^{(0)}} \partial_+ (\gamma_{(0)}^{ab} \gamma^{(1)}_{ab})\bigg\vert_{\tilde A}.
\end{split}
\end{align}

As a warm-up consider a ball-shaped entangling surface with a corresponding extremal surface at $r^+ = \rho^+_0, r^- = -\rho^+_0$ placed in a planar black hole background,
\begin{align}
ds^2 = \frac{1}{z^2}\left( - (1- \mu z^d ) dt^2 + \frac{dz^2}{(1-\mu z^d)} + dx^2 \right),
\end{align}
at leading order in $\mu$. The equations for the extremal surface now become at first order
\begin{align}
\label{eq:Diffeo}
0 =  &\left. \frac 1 2 \sqrt{\gamma^{(0)}}\partial_\pm \left( \gamma_{(0)}^{ab} \gamma^{(1)}_{ab} \right) -  \partial_a\left(\sqrt{\gamma^{(0)}} \gamma_{(0)}^{ab}  g^{(1)}_{\pm b}\right) \right|_{\tilde A}.
\end{align}
We can use the symmetry of the perturbation under time translations and regularity at the boundary to find a vector field $v$ that generates a diffeomorphism $g \to \mathcal L_v g$ which locates the extremal surface in the perturbed geometry at the same coordinate location as the extremal surface in the unperturbed geometry.
\begin{align}
\label{eq:HW1}
v_+ &= -\frac \mu{64}\sin \theta ( 1 + \sin^2 \theta) (r^+ - r^-)^2,\\
\label{eq:HW2}
v_- &= \frac \mu{64}\sin \theta ( 1 + \sin^2 \theta) (r^+ - r^-)^2,\\
\label{eq:HW3}
v_\theta &= \frac \mu {64} (r^+ - r^-)^3 \cos^3 \theta,\\
\label{eq:HW4}
v_\phi &= 0.
\end{align}
This diffeomorphism brings the metric perturbation into the form
\begin{align}
\begin{split}
\delta ds^2 = &\frac \mu {8} (r^+ -r^-) \frac{1 + \sin^2 \theta}{\sin \theta} dy^+dy^- + \frac \mu {32} (r^+ -r^-)^3 \cos \theta \cot \theta d\theta^2 \\& - \frac \mu {32} (r^+ -r^-)^3 \cos^3 \theta \cot \theta d\phi^2.
\end{split}
\end{align}
The only non-vanishing components of the metric in the new coordinates are $g_{+-}, g_{\theta\theta}$ and $g_{\phi\phi}$. In particular, we have that $g^{(1)}_{-a} = 0 = g^{(1)}_{--}$. The main benefit of these coordinates is that equation \eqref{eq:Diffeo} holds automatically. Hence at least for a ball-shaped entangling surface we are in Hollands-Wald gauge and the extremal surface is located at $r^\pm = \pm \rho^+_0$. It can be seen from the metric that lines of constant $r^\pm$ are lightlike and therefore we know that the new entangling surface still is on the bulk lightcone of a point $p$ at the boundary.

From this we can conclude that the entanglement wedge associated to any region bounded by a lightcone does not contain any point outside the causal wedge. As we have seen this is true for ball-shaped regions. A deformation of the entangling surface cannot change this, since the boundary domain of dependence is smaller than that of some ball-shaped region. At the same time, the extremal surface cannot lie within the causal domain of dependence and therefore we must conclude that the extremal surface also lies on the lightcone. 

This means that the transformations \eqref{eq:HW1} -- \eqref{eq:HW4} bring the RT surface to its correction $r^+$ location. The only additional adjustment we need to make to the coordinate system is to reparameterize $r^-$ around the extremal surface, e.g.~by rescaling the $r^-$ coordinate in an angle-dependent way.

To find a solution to the general Hollands-Wald gauge condition, equation \eqref{eq:HWGaugeCondition}, we alter the plus-component of the vector field, $v_+ \to v_+ + \tilde v_+(\theta,\phi)$, around the extremal surface such that it shifts the extremal surface into its new correct location on the lightcone. This vector field can be chosen such that at the extremal surface $\tilde A$ it remains constant along $r^-$ and $r^+$ and thus depends only on $\theta$ and $\phi$. It should be clear that such a solution exists, since at the boundary the correction $\tilde v_+(\theta,\phi)$ vanishes and is smooth everywhere else. More formally, in this case equation \eqref{eq:HWGaugeCondition} reduces to
\begin{align}
\begin{split}
&\frac \mu {16} \partial_\theta (\cot \theta \partial_\theta (\rho_0^+-\Lambda(\theta,\phi))^2  ) + \frac \mu {16}  \tan \theta \partial^2_\phi (R-\Lambda(\theta,\phi))^2\biggr\vert_{\tilde A} \\ ={ } &\partial_\theta(\cos \theta (\partial_\theta \tilde v_+(\theta,\phi) + 2 \cot \theta \tilde v_+(\theta,\phi))) +\frac{1}{\cos \theta} \partial^2_\phi \tilde v_+(\theta,\phi)\biggr\vert_{\tilde A}.
\end{split}
\end{align}
For small deformations of the ball shaped entangling surface we can write $\Lambda(\theta,\phi)$ as a series expansion in the deformations. At first order, this gives us a linear PDE which can be solved. Higher orders become inherently non-linear and thus this equation is in general very hard to solve. An interesting observation one can make for small $n=1$ deformations of the entangling surface is that the linear order correction is zero.

\bibliographystyle{JHEP}
\bibliography{Cone}

\end{document}